\documentclass[aps,pra,twocolumn,groupedaddress, floatfix, raggedbottom]{revtex4-1}
\usepackage{amsmath}
\usepackage{url}
\usepackage{graphicx}
\usepackage{hyperref}
\usepackage{xcolor}
\hypersetup{
    colorlinks,
    linkcolor={red!50!black},
    citecolor={blue!50!black},
    urlcolor={blue!80!black}
}

\begin{document}

\title{Developing Quantum Annealer Driven Data Discovery}

\author{Joseph Dulny III}
\email[]{dulnyiii\textunderscore joseph@bah.com}
\author{Michael Kim}
\email[]{kim\textunderscore michael3@bah.com}
\affiliation{Booz Allen Hamilton\\308 Sentinel Drive, Suite 100\\Annapolis Junction, MD 20701 USA}

\date{\today}

\begin{abstract}
Machine learning applications are limited by computational power. In this paper, we gain novel insights into the application of quantum annealing (QA) to machine learning (ML) through experiments in natural language processing (NLP), seizure prediction, and linear separability testing. These experiments are performed on QA simulators and early-stage commercial QA hardware and compared to an unprecedented number of traditional ML techniques. We extend QBoost, an early implementation of a binary classifier that utilizes a quantum annealer, via resampling and ensembling of predicted probabilities to produce a more robust class estimator. To determine the strengths and weaknesses of this approach, resampled QBoost (RQBoost) is tested across several datasets and compared to QBoost and traditional ML. We show and explain how QBoost in combination with a commercial QA device are unable to perfectly separate binary class data which is linearly separable via logistic regression with shrinkage. We further explore the performance of RQBoost in the space of NLP and seizure prediction and find QA-enabled ML using QBoost and RQBoost is outperformed by traditional techniques. Additionally, we provide a detailed discussion of algorithmic constraints and trade-offs imposed by the use of this QA hardware. Through these experiments, we provide unique insights into the state of quantum ML via boosting and the use of quantum annealing hardware that are valuable to institutions interested in applying QA to problems in ML and beyond.
\end{abstract}


\maketitle

\section{Introduction and Background}
Quantum computation is an area of great interest to a variety of fields, including, but not limited to: cryptography/cryptanalysis~\cite{shor1994algorithms, shor1997polynomial, boneh1995quantum, bernstein2009post, bennett1984quantum}, combinatorial optimization~\cite{han2000genetic, han2002quantum}, protein folding~\cite{perdomo2012finding}, genetic programming~\cite{spector1999quantum}, and machine learning~\cite{aimeur2006machine, sasaki2002quantum, pudenz2013quantum, denchev2013binary}. In theory, quantum computers can outperform classical, transistor-based computers significantly in certain areas. For this reason and the applications described above, they are an area of heavy research across the fields of physics, computer science, electrical engineering, and mathematics.

In this work we discuss calculations run on a commercial quantum annealing machine. The computational core of this hardware is a 512-point array of superconducting quantum interference devices (SQUIDs) that function as the qubits of the system. The use of SQUIDs as qubits has been studied in the past~\cite{chiorescu2003coherent, chiorescu2004coherent, yang2003possible, zhou2002quantum}. In practice, all 512 of the qubits are not functional when the chip is finished, and so the user must adapt to the network of functional qubits present in the system. The system used in these experiments had 476 functional qubits. Additionally, due to the hardware graph of the qubits, each qubit in the system is not coupled with every other qubit. This means that coupling between every given pair of qubits in the system is not possible and that the user must adapt and program the system such that all desired couplings are possible within the connectivity graph of the chip.

Quantum annealing~\cite{finnila1994quantum, kadowaki1998quantum, brooke1999quantum, farhi2000quantum, farhi2001quantum, santoro2002theory, santoro2006optimization} is a fundamentally different process from the algorithmic approach that would be implemented on quantum gate computers. Instead of functioning as a universal quantum computer, a quantum annealer has been shown to be useful in solving specific problems, namely NP-hard combinatorial optimization problems~\cite{farhi2001quantum, garey2002computers}. In quantum annealing, the goal is to find a global minimum of a given function by manipulating the Hamiltonian of a system of qubits. This is done by programming the qubits to have a two-term Hamiltonian: an ``easy'' term whose ground state is known, and a ``problem'' term whose ground state is the solution to the problem the user wants to solve. This problem is always the minimization of a given objective function. The system is then initialized into the ``easy'' portion of the Hamiltonian and put into its ground state. Then the system is very slowly evolved into the ``problem'' portion of the Hamiltonian and, via the adiabatic theorem~\cite{born1928beweis}, the system should stay in its ground state, and thus display the desired solution. More formally, the quantum annealer solves the problem of minimizing the Hamiltonian
\begin{equation}
H\left (\pmb{\sigma}\right ) = \sum_{i,j}J_{ij}\sigma_i\sigma_j +\sum_{i}h_{i}\sigma_i
\end{equation}
where $\pmb{J}$ represents interactions between a given pair of variables, each $\sigma_i \in\{0,1\}$ represents a binary variable, and $\pmb{h}$ represents an external bias being applied to each variable. The goal is to set the values of $\pmb{\sigma}$ such that the Hamiltonian is minimized. This problem class is known as quadratic unconstrained binary optimization (QUBO). To solve this problem, the quantum hardware sets up a two-term Hamiltonian
\begin{equation}
H \left (\pmb{\sigma}, \xi \right ) = \left ( 1-\xi \right )H_e+\xi H_p
\end{equation}
for $0\le\xi\le1$. Here $H_e$ is the initial ``easy'' Hamiltonian that has an easily locatable global minimum and $H_p$ is the ``problem'' Hamiltonian that is to be minimized. As $\xi$ is slowly changed from 0 to 1, an ideal system will adiabatically evolve from the ground state of $H_e$ to that of $H_p$ and the solution will be found.

This method stands in contrast to thermally driven simulated annealing~\cite{aarts1988simulated, kirkpatrick1984optimization}. In simulated annealing, a global minimum is searched for using thermal fluctuations and gradual temperature reduction. This method can lead to a local rather than global minimum. Quantum annealing, however, makes use of quantum properties such as tunneling and the adiabatic theorem, allowing the system to evolve toward its ground state. This evolution is guaranteed to reach the ground state in the ideal, theoretical case of zero temperature and very long annealing time, but in practice non-zero temperatures and other factors could cause excitations that lead to suboptimal solutions.

There have been mixed views within the scientific community as to whether the hardware studied here is actually exhibiting quantum behavior or speed-up~\cite{van2007quantum, cho2011controversial, johnson2011quantum, mcgeoch2013experimental, vinci2014distinguishing, lanting2014entanglement, ronnow2014defining, denchev2015computational, benedetti2015estimation, boixo2016computational}. Several studies have concluded that entanglement and coherence were observed in the units they tested. The authors of one study \cite{lanting2014entanglement} state that this provides ``an encouraging sign that QA is a viable technology for large-scale quantum computing.'' A June 2014 study~\cite{ronnow2014defining} concluded that the unit they experimented with showed no signs of quantum speedup when compared with classical methods on the specific problems they tested. However in this same work the authors note that ``Our results do not rule out the possibility of speedup for other classes of problems and illustrate the subtle nature of the quantum speedup question.'' More recent results from December 2015 \cite{denchev2015computational} show a speedup on certain problem cases when compared with specific classical algorithms, but not in general. In light of the academic debate and scrutiny surrounding this hardware, experimental QA calculations and their comparisons with classical computers and traditional methods and models are of great scientific interest.

The primary focus of this study is the development, application, and evaluation of quantum annealing enabled machine learning. Since this field contains many NP-hard combinatorial optimization problems, quantum annealing lends itself naturally to applications in it~\cite{kaminsky2004scalable, neven2009training, mukherjee2014multivariable, babbush2014construction, o2014bayesian, adachi2015application}. Despite the existence of theoretical studies on the subject for some time, the actual experimental application of quantum annealers to machine learning problems is an extremely recent innovation. Due to the availability of this quantum annealing hardware, several experimental studies on machine learning and performance have now been completed. A 2009 study by Neven et al.~\cite{neven2009nips} focused on using this type of quantum annealing hardware for binary classification by developing the QBoost algorithm. In this algorithm, the quantum annealing hardware functions as an oracle to optimize the loss function generated during each iteration. In this study, they found that the annealer and algorithm they used outperformed a limited set of classical software classifiers. 

Within machine learning, we specifically focused this study on quantum annealing enabled binary classification via regularized boosting. The goal of these techniques is to select the best ensemble of weak classifiers that, when assembled together in a voting quorum, form the most accurate strong classifier. 

The current difficulty with execution of boosting on this system is that the number of qubits available (476 for this study) is small, and so optimizing a large set of weak classifiers in one shot is not possible. Neven et al.\ addressed this in their QBoost algorithm~\cite{neven2009nips, neven2012qboost}. QBoost is a modified version of AdaBoost that iteratively loops through a large pool of weak classifiers and optimizes subsets of them. It then greedily ensembles one or more weak classifiers into the strong classifier if their addition improves the classification accuracy over the validation set. This greedy approach may not find the globally optimal strong classifier due to the promotion of a weak classifier that is not a member of the best possible subset of weak classifiers. 

In this work we present and test our resampled improvement to QBoost (RQBoost), against a large number of traditional machine learning techniques on a variety of datasets. Prior evaluations of quantum annealer enabled machine learning only tested against a limited number of traditional algorithms, and so these results give a much clearer picture of where these techniques stand in comparison to modern machine learning on classical hardware. 

\section{General Methods}

\subsection{Overview}

Three different datasets were studied and had binary classifiers constructed for them using both classical and quantum techniques. Each dataset had candidate weak binary classifier sets evaluated and ensembled using QBoost/RQBoost and either quantum annealing hardware or a quantum annealing simulator. The performance of these strong classifiers was compared to strong classifiers formed using traditional machine learning techniques, such as logistic regression with L1 and L2 regularization, gradient boosting machines, and random forests. Unlike initial benchmarks of QBoost~\cite{neven2009nips, neven2012qboost}, which focused on comparisons to AdaBoost, we compared the performance of QBoost and RQBoost to a wide range of traditional, state of the art machine learning algorithms. This allows for a broader view of the current state of quantum machine learning via boosting in comparison to traditional machine learning. 

\subsection{The RQBoost Algorithm}
\label{rqboost}
RQBoost is our new binary classifier algorithm that extends the original QBoost through resampling and outputting probability estimates. As with QBoost, RQBoost uses quantum annealing hardware as an oracle for minimization of the loss function constructed during each iteration. The loss function minimized is of the form
\begingroup\makeatletter\def\f@size{9}\check@mathfonts
\begin{equation*}
H(\boldsymbol{w})=\frac{1}{2}\sum_{s=1}^{S}\left [ \frac{1}{T_{out}+Q} \sum_{j=1}^{Q}w_j F_j(x_s) - \hat{y}_s \right ]^2 + \lambda\sum_{j=1}^{Q}w_j
\end{equation*}
\endgroup
where
\begingroup\makeatletter\def\f@size{9}\check@mathfonts
\begin{equation*}
\hat{y}=y_s -\frac{1}{T_{out}+Q} \sum_{t=1}^{T_{out}}F_t(x_s) .
\end{equation*}
\endgroup
Here $w_j \in \{0,1\}$ are the weights of the weak classifiers, $S$ is size of the training set, $T_{out}$ is the size of the current strong classifier, $Q$ is a tunable parameter typically set to the number of variables that an optimization run can handle, $F_j$ represents the weak classifiers currently under consideration, $F_t$ represents the weak classifiers that have already been accepted into the strong classifier, $\lambda$ is the real-valued regularization strength, and $y_s \in \{-1,1\}$ is the label of a given training item.

The original QBoost algorithm only uses a fraction of the data as a result of requiring a train, validation, and test split of the input data. This lowers the power of the estimator since not all available data is leveraged explicitly in the training aspect of the algorithm. Repeatedly resampling and then producing three partitions of the data mitigates this problem, and thus is the key improvement offered by RQBoost. As the number of resamples increases to infinity, the probability of including all of the data in the training split at least once goes to one. Thus, given infinite resamples, all data is explicitly used to train the algorithm. In the practical case of many (but not infinite) resamples, the data is much more broadly used than in the case of a single split, as is used by QBoost.

In addition, the process of resampling and generating many individual strong classifiers allows RQBoost to produce probability estimates for each class instead of outputting binary class estimates by majority vote, as is the case with binary-weighted QBoost. In each resample iteration of RQBoost, a new strong classifier is generated. For a given test, the fraction of votes for a particular class the weak ensemble classifiers cast in a given strong classifier is calculated. The normalization function then divides the fractional votes for a particular class by the total number of strong classifiers produced by RQBoost. The average of these scores across a large number of repeated resampled runs produces a final probability estimate of class.

\subsection{Usage Specifics}
\label{usageSpecifics}
During the execution of these algorithms, QUBO problems are generated via specification of the $h$ and $J$ coefficients. These QUBO problems represent minimizations of the loss function used in QBoost/RQBoost. These problems are submitted to the quantum hardware, and the minimum energy solution of the returned solutions is used in the continued execution of the algorithms (see~\cite{neven2012qboost} for details). These problems are often fully connected, meaning that there is an interaction term $J_{ab} \sigma_a \sigma_b$ for every pair of variables $\sigma_a$ and $\sigma_b$ in the problem. 

\begin{figure}
\centering
\includegraphics[width=\columnwidth]{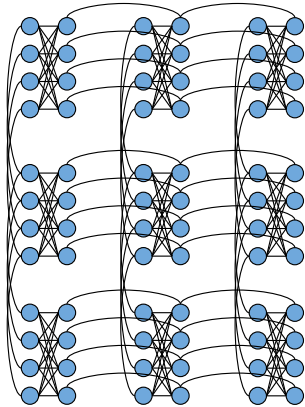}
\caption{Nine unit cells of the graph structure used for the qubits in the QA hardware used in this study. Circles represent flux qubits which can be biased with $h$ values and lines represent couplings between qubits which are biased with the $J$ values in a given problem. The above structure is tiled to create the the graph of the entire chip.}
\label{hardwareGraph}
\end{figure}

In order to solve these using this type of quantum annealer, the problem must be embedded into the hardware qubit graph structure. The hardware qubit graph, a piece of which is shown in schematic form in Figure~\ref{hardwareGraph}, is a regular repeating structure of unit cells consisting of 8 qubits. The actual connectivity varies from chip to chip due to nonfunctional qubits and connectors. The highest connectivity possible in this configuration is 6 for the bulk and 5 for the edges, so in order to embed a fully connected problem into the hardware, qubits must be chained together to form logical qubits with higher connectivity. Finding an embedding that optimally satisfies the problem couplings and fits into the specific hardware graph with minimal chaining is a nontrivial problem that is computationally NP-hard and therefore is typically solved via heuristic methods. The use of heuristic solvers to find embeddings implies that the usage of suboptimal embeddings to solve more fully connected problems is both commonplace and necessary.

The process of actually using the hardware to solve a problem that involves qubit chains involves several extra steps compared to solving a problem that fits into the native hardware format. The chain strength of the qubit chains corresponds to the artificial $J$ values these couplings are given. For a given problem, several different chain strengths must be tried due to the following trade-off: if the chain strengths are set too low, the annealing process will not preserve the larger logical qubits, and therefore will return answers that do not maintain the problem structure. Conversely, if the chain strengths are set too high, then the relative strength of the $J$ coefficients corresponding to logical qubit connections will be very large compared to the actual $h$ and $J$ values of the problem submitted, and the anneal will strictly preserve the logical qubit chains without actually finding the minimum to the problem Hamiltonian. There are cases where acceptable values of chain strength for the problems being considered can be determined in advance. These procedures and trade-offs must be taken into account when using this quantum hardware to solve non-natively structured problems. Thus it can be seen that using the quantum hardware to solve non-natively structured QUBO problems is a nontrivial process.

\section{Name Classification}
\label{names}

\subsection{Methods}
\label{namesMethods}
As an initial study, the performance of QBoost, RQBoost and R's implementation of random forest were compared on the names corpus v1.3 from the Natural Language Toolkit~\cite{namescorpus} with the area under the receiver operating characteristic curve (AUC) as the evaluation metric. The data used was a labeled set of 2943 male names and 5001 female names. The classifiers used for this dataset were very simple: if a name ended in a given letter, then it was classified as male or female. Each combination of letter and male/female was created, resulting in a pool of 52 weak classifiers.

The rows in this names dataset were shuffled and prepared for 10-fold cross-validation. This resulted in 10 random horizontal partitions of the dataset. 10 train and test set pairs were constructed from partitioned dataset, with each pair containing all 10 folds. 9 of the folds in a given pair were assigned to the training set, and the remaining fold was the test set. This resulted in a distribution of 10 AUC values for each algorithm tested. Here, we examine these empirical distributions as a means to compare the performance of QBoost, RQBoost and R's implementation of random forest with default parameters.

In these experiments, we compared the performance of QBoost and RQBoost using a hardware graph structured brute force software solver rather than the quantum hardware itself. This brute force solver had a built-in qubit structure that was in the same pattern as the real quantum hardware and thus required the heuristic embedding finder and iterative solving by varying the chain strength to find a suitable value. Here, we deliberately used a brute force software solver to isolate the algorithm from the experimental noise which can occur with true quantum hardware runs. This allowed a more direct comparison of the algorithms' performance.

\subsection{Results}

\begin{figure}
\centering
\includegraphics[width=\columnwidth]{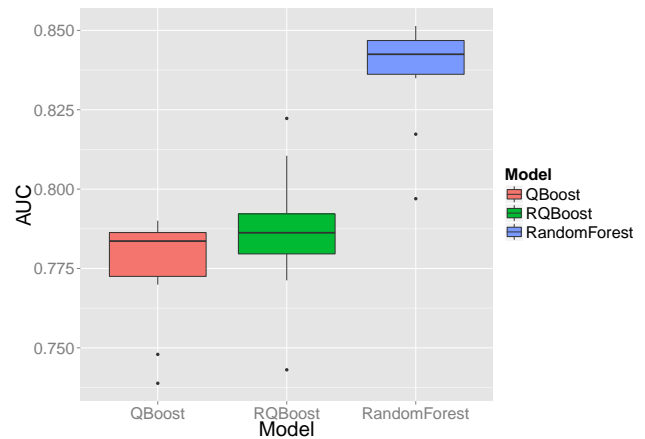}
\caption{Performance comparison of QBoost, RQBoost, and the R's implementation of random forest. This plot shows data from Table~\ref{namesTable}. All values are area under the receiver operating characteristic curve (AUC).}
\label{namesFigure}
\end{figure}

The R implementation of random forest with default parameters and random seed set to one was used to produce one empirical AUC value for each of the train/test pairs. The QBoost and RQBoost algorithms were used on each of the train/test pairs. The distributions of the AUC values obtained are given in Table~\ref{namesTable} and plotted in Figure~\ref{namesFigure}.

\begin{table*}
\centering
\caption{Distribution of the empirical AUC values obtained for the names data using three different machine learning techniques: R's random forest, QBoost, and RQBoost.}
\vspace{6pt}
\begin{tabular}{llllllll}
\hline
Technique     & Min.  & 1st Qu. & Median & Mean  & 3rd Qu. & Max.  & Std. Dev. \\ \hline
Random Forest & .7970 & .8362   & .8425  & .8367 & .8468   & .8514 & .01697    \\
QBoost        & .7388 & .7725   & .7836  & .7753 & .7863   & .7900 & .01785    \\
RQBoost       & .7431 & .7796   & .7863  & .7861 & .7922   & .8223 & .02137
\end{tabular}
\label{namesTable}
\end{table*}

RQBoost outperformed QBoost in every summary statistic from minimum to maximum. In this particular dataset, RQBoost performed only 30 resamples but still made a clear improvement over the AUC scores of QBoost. With so few resamples, it is promising to see a significant performance improvement like this. It is expected RQBoost's performance will continue to increase with more resamples. As a comparison to traditional machine learning techniques, the empirical AUC distribution produced by R's random forest was noticeably better than both QBoost and RQBoost. This was expected, as this is a highly refined technique which is used frequently and does not face the same restrictions as QBoost and RQBoost.

\section{Case Study: Seizure Prediction}

As a much more complicated follow-up to our initial study of name classification, we conducted the first ever application of quantum hardware to seizure prediction in Kaggle's American Epilepsy Society Seizure Prediction Challenge~\cite{kaggleseizure}. The objective of the challenge was to estimate the probability of a seizure given EEG-based time series data in humans and canines. For those suffering with epilepsy, having foreknowledge of an oncoming seizure would be extremely beneficial and could even be life-saving. On top of performance benchmarking, this case study demonstrates important constraints when programming for a quantum annealer versus a typical classical computer.

\subsection{Methods}

Intracranial EEG clips from humans and canines with epilepsy were provided by the online data science competition website Kaggle~\cite{kaggleseizure}, who sourced them from the NIH-sponsored International Epilepsy Electrophysiology Portal~\cite{ieeg}. The sample rate for this data was 5000 Hz, and various numbers of electrodes recording the voltage outside particular locations on the subjects' brains were used. Datasets were all 10 minutes in length and divided into two categories: \textit{interictal}, which is normal brain function temporally distant from a seizure; and \textit{preictal}, which is brain activity that came one hour before a seizure. The interictal data was chosen to be at least one week before and after seizure events for canines and at least four hours before and after seizure events in humans.

To run QBoost, RQBoost, and traditional machine learning algorithms on this data, features were extracted using the methods described in Appendix~\ref{seizureFeatures}. However, traditional machine learning algorithms, such as linear regression or decision trees, use real-valued features where quantum techniques like QBoost and RQBoost are limited to binary-valued features. Given a very large number of qubits to work with, this would not be a problem because of techniques such as the hash trick or one-hot encoding. Despite this, the largest available quantum annealers only contain approximately 1000 qubits with sparse connectivity, which is not yet large enough to one-hot encode real-valued features. Thus it is not possible to use QBoost/RQBoost in exactly the same manner as typical machine learning algorithms.

Furthermore, due to imperfect transmission to the system, the precision with which weight coefficients $h$ and couplings $J$ can be defined on the hardware within one standard deviation is approximately $\pm 0.05$ of the full range~\cite{precision}. This will be discussed further below in Section~\ref{lessonslearned}, but this means that at this time not all problems can be expressed perfectly on the hardware. Thus we do not directly compare traditional machine learning to quantum machine learning. Instead we show very specific differences between quantum and traditional methods graphically across a large range of trials on the seizure dataset.

Given these difficulties presented by comparison and several more elaborated on in \ref{lessonslearned}, Lessons Learned, we have opted here for the approach outlined below.

\subsection{Results}

\begin{figure}
\centering
\includegraphics[width=\columnwidth]{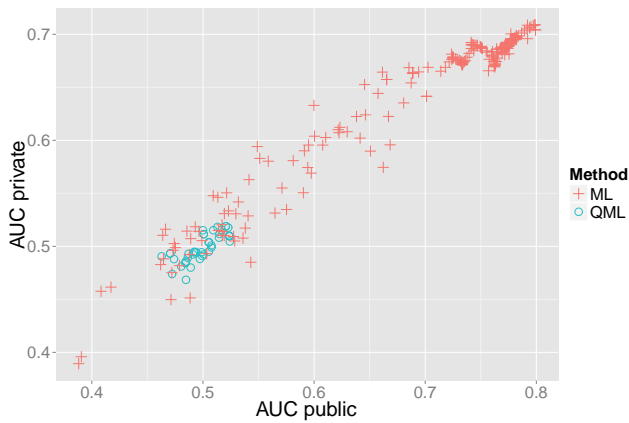}
\caption{Aggregate performance comparison of quantum machine learning (QML) and traditional machine learning (ML) techniques. QML techniques included RQBoost and QBoost and were executed on quantum hardware, while ML techniques included a variety of random forests, gradient boosting machines, and penalized regression. The QML methods here had access to a smaller set of features than the ML methods.}
\label{seizureFigure}
\end{figure}

Results are represented in Figure~\ref{seizureFigure}, which is a plot of aggregate AUC data for all quantum machine learning (QML) and traditional machine learning (ML) techniques used. Kaggle functions by having a public and private test set. Due to the timing of these experiments, the traditional machine learning algorithms were executed before the competition closed and all quantum machine learning testing was done after the challenge concluded. In general, RQBoost and QBoost had lower mean AUC scores than linear regression, gradient boosting, or random forest techniques when given the same binary features. However, due to qubit limitations, RQBoost and QBoost were only given limited testing across a subset of features versus the multiple trials used by traditional machine learning techniques. Conversely, the chosen ML algorithms were applied to large datasets involving real-valued features. Using these ML methods, we were able to finish 58th out of 504 entrants on the challenge with a private leaderboard AUC score of 0.70419. This relatively high placement speaks to the quality of the feature extraction we performed and legitimizes their use in testing QBoost and RQBoost. Due to the various limitations imposed by these algorithms and the quantum hardware, we do not believe it is currently possible to achieve such a high score using only quantum annealer based machine learning. This is due to several factors, including the fully connected loss functions generated by QBoost/RQBoost and various properties of the hardware that will be outlined below in Section~\ref{lessonslearned}. As a result, we were motivated to investigate RQBoost and the quantum annealer in a much more controlled setting, which is described in the next section.

\section{Linearly Separable Data}

\subsection{Methods}

To further evaluate QBoost and the quantum hardware in a more controlled setting, a computer-generated dataset was used to test the quantum annealer/QBoost's ability to separate two classes perfectly via a hyperplane. This dataset contained weak classifiers such that a subset of the classifiers, when ensembled, perfectly classified all of the data. Individually, none of the classifiers in the perfect ensemble were highly correlated with the ground truth of the dataset. Additionally, weak classifiers that were highly correlated, albeit imperfectly, with the ground truth of the dataset were included as ``bait'' for the boosting algorithm. The goal in studying this dataset was to see if the quantum annealer/QBoost or traditional machine learning techniques could separate the perfect ensemble of weak classifiers and not take the ``bait.''

\begin{figure}
\centering
\includegraphics[width=\columnwidth]{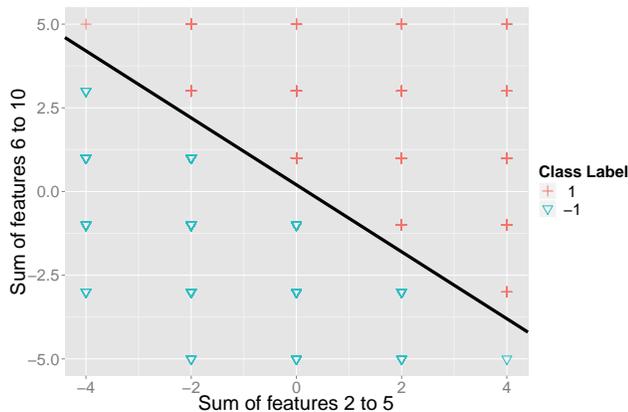}
\caption{Graphical representation of the linearly separable data used.}
\label{linsepFigure}
\end{figure}

This data was generated by creating a random binary matrix of -1 and 1 with 1000 rows and 13 columns. After this matrix was generated, the first column was set to be the ground truth by combining columns 2 through 10 inclusively. The sum of the sign of columns 2 through 10 inclusive replaced column 1 and served as ground truth. Column 13 was set as the highly correlated ``bait'' column, where 91\% of the column correctly predicted the ground truth column 1.  The data is represented graphically in Figure~\ref{linsepFigure}.

\begin{table}
\centering
\caption{The confusion matrix of the linearly separable data.}
\begin{tabular}{l|l|c|c|c}
\multicolumn{2}{c}{}&\multicolumn{2}{c}{Predicted}&\\
\cline{3-4}
\multicolumn{2}{c|}{}&Positive&Negative&\multicolumn{1}{c}{Total}\\
\cline{2-4}
Ground & Positive & $459$ & $47$ & $506$\\
\cline{2-4}
Truth & Negative & $43$ & $451$ & $494$\\
\cline{2-4}
\multicolumn{1}{c}{} & \multicolumn{1}{c}{Total} & \multicolumn{1}{c}{$502$} & \multicolumn{1}{c}{$498$} & \multicolumn{1}{c}{$1000$}\\
\end{tabular}
\label{confusionmatrix}
\end{table}

\subsection{Results}

Since the generated data is linearly separable by combining columns 2 through 10 and taking the sign, we assumed the ideal behavior of the quantum hardware/QBoost would be separation of the perfect ensemble of classifiers. However, this did not turn out to be the case. The quantum annealer/QBoost repeatedly included the bait column 13 along with columns 2 through 10 inclusive. This led to an imperfect classifier. The confusion matrix is shown in Table~\ref{confusionmatrix}. This test was conducted for a wide range of regularization and chain strength values and was never observed to perfectly separate the generated dataset. Traditional machine learning methods such as R's glmnet were able to separate the perfect ensemble. Additionally, regularized regression with logistic loss also succeeded for a variety of regularization strengths. This is expected behavior for both LASSO and ridge regression. Resampling (RQBoost) was not used for this dataset. 

Upon investigation of these results we determined why QBoost and the commercial quantum annealer were unable to perfectly separate this dataset. While the problem of linear separability may be a trivial one for most machine learning algorithms, the structure of QBoost forces some extra constraints to be placed on the structure of the loss function that account for the observed results. These constraints are a result of the specialized structure of the quantum hardware. The hardware optimizes a loss function of the form $(\hat{y} - label)^2$, where $\hat{y}$ is the predicted value and $label$ is the ground truth. The full form  of this loss function was given in Section~\ref{rqboost}. The additional constraint is that $\hat{y}$ is limited to the class $\hat{y} =Q^{-1}\sum_iV_i(\mathbf{x})$, where $V_i$ is the vote of the \textit{i}\textsuperscript{th} weak classifier, $\mathbf{x}$ is the input pattern, and $Q$ is a tunable parameter equal to the number of features in the current implementation. Due to this, there are cases where, even if the hardware found the global minimum of the above loss function, this solution would not correspond to the solution of the linearly separable problem.

If $Q$ is too large, the optimal solution returned will fail the separability test regardless of regularization choice. This sensitivity to $Q$ indicates that QBoost may not be well-suited for all problems. 

\section{Lessons Learned}
\label{lessonslearned}

Over the course of this project, we learned many details about the use and limitations of these quantum annealers. In this section on our lessons learned, we would like to briefly elaborate on them for the benefit of the industrial and academic community.

One of the primary considerations when using this hardware is the graph structure of the qubits themselves. A defect-free segment of the qubit graph was shown in Figure~\ref{hardwareGraph}. The connectivity and uniqueness (due to defects) of this repeating graph must be taken into account when considering problems to submit to the hardware. In the best case, most qubits have connections to only 6 other qubits, which imposes restrictions on the problem structure that can be solved.

Heuristic methods for embedding more highly connected problems into the quantum hardware exist, but their use comes with several costs. First, a penalty in qubits must be paid as multiple qubits are chained together with large $J$ values to form logical qubits with higher order connectivities. In the worst-case of having fully connected variables in the problem Hamiltonian, this qubit penalty scales approximately as the square of the number of variables. For example, the 476 variable chip we used in these studies could have a maximum of 25 fully connected variables. This penalty must be carefully considered during problem selection.

Another penalty coming with highly connected problems is the process of chaining qubits to form an embedding. As described in Section~\ref{usageSpecifics}, each problem submission using chaining might have to be submitted several times in order to find the optimal value for chain strength. These extra submissions are costly and should be taken into consideration when planning calculations using embeddings. In some cases, a chain strength can be found that is likely to succeed for most problems in an embedding graph and so the need for multiple submissions for every problem is reduced.

In summary, for a non-natively structured problem, there are several considerations that must be made during the problem-solving process. One must first heuristically search for an embedding, pay the cost in additional qubits to implement the embedding, and possibly perform multiple calculations for different values of chain strength as problems are submitted. Thus the process of solving non-natively structured problems can be costly. These combined costs highly motivate the user to structure problems in the native hardware format if possible.

Another crucial hardware consideration is the precision with which the hardware can physically map input $h$ and $J$ values to the hardware. This arises as a combination of systematic, random, persistent, and transient errors in the system and is known collectively as intrinsic control error (ICE). The precision with which $h$ and $J$ can be defined on the hardware within one standard deviation is approximately $\pm .05$, which is approximately 5\% of full range of possible values. This is an additive error with a mean of zero~\cite{precision}. This is important to note, as it is quite easy to over-specify the input coefficients when working in 32-bit or 64-bit variables commonly found in classical computer programming languages. When presented with a problem that over-specifies the input coefficients, the hardware will truncate them to lower precision values. Due to the combinations of ICE and hardware truncation, high precision problems are represented imperfectly on the quantum hardware. This implies that a problem solved on the hardware may be different from the problem that was submitted to it~\cite{theking}.  

It is also important to understand that there is never a guarantee that the quantum annealing hardware will obtain an optimal solution for a submitted problem~\cite{theking}. This is due to the fact that any real quantum annealing system will be unavoidably coupled to its surrounding environment and therefore may experience interactions with the environment that disrupt the annealing process. 

Concerning QBoost and RQBoost, we note that the structure of the loss function used in these algorithms tends to generate fully connected QUBO problems to submit to the hardware. We have observed that this non-native structure imposes penalties that would best be avoided altogether if possible. Given sparsely connected QUBO problems, many more variables could be considered at once, and the extra steps required to embed highly connected problems into the hardware graph could be reduced or avoided altogether. Since the number of qubits required to represent a fully connected problem in the hardware graph scales approximately as the square of number of variables used, we feel this reduction in connectivity is especially important.

\section{Conclusion}

In this work, we have used classical and quantum hardware to perform machine learning experiments on natural language, seizure EEG, and linearly separable datasets. We extended and improved the QBoost algorithm via resampling and we call this result RQBoost. RQBoost outperformed QBoost on the datasets tested here using a 476 qubit commercial quantum annealer, but both quantum algorithms were outperformed by current machine learning techniques. These results are important because they show how current quantum machine learning via boosting compares to a variety of state-of-the-art traditional machine learning techniques, which was previously unknown. 

These experiments also provide valuable insights into the usage of quantum annealing hardware and algorithms and their applicability to machine learning. Since this is an emerging field, we addressed specific considerations that users must make before implementing a quantum annealer based solution. These included the need to account for intrinsic control error in programming $h$ and $J$ coefficients, to structure the problem as closely to the native hardware format as possible, and to embed non-natively structured problems properly to ensure that global optimums are more likely to be obtained. Despite these considerations, there is no guarantee that global optimums will be returned. Additionally, the details and limitations of QBoost and RQBoost as applied with quantum annealing hardware were discussed at length, primarily the tendency of the these algorithms to generate fully connected QUBO problems. Fully connected problems require extra steps to solve, and these steps reduce the number of variables that can be optimized at once and therefore less effectively leverage the quantum hardware.

Studying QBoost and RQBoost allowed us to more accurately define the specific requirements of machine learning algorithms designed for execution on quantum annealing hardware. These results motivate the design and testing of the next generation of purpose-built quantum machine learning algorithms.

\appendix
\section{Seizure Data Engineering}
\label{seizureFeatures}

The seizure dataset consists of Intracranial EEG (iEEG) time series data with multiple channels corresponding to electrode locations in both humans and canines. This training data is labeled as ``preictal'' for pre-seizure data segments, or ``interictal'' for non-seizure data segments. Given this real-valued time series data, we obtained a feature matrix (with rows as observations and columns as features) through a variety of means. We produced many different models (to ensemble), each based upon a different combination of transformations. However, the general outline of transformation is as follows:
\begin{enumerate}
\item Use a filtering method such as a median filter or Gaussian filter in the time domain. This smooths out the data and removes outliers.
\item Pick a window or multiple overlapping windows which will be used in Step 3.
\item Given this filtered data, make summary statistics of the data within the window or windows obtained in Step 2. These summary statistics can include the mean, median, max, time index of max, min, time index of min, standard deviation, higher order moments, and quantiles. This method can also be applied to frequency domain data after a discrete fast Fourier transform is applied.
\item If using multiple channels, use Pearson correlation to create a correlation matrix. Take the lower triangle of this correlation matrix as the final features. Otherwise, just take a subset of channels from which to calculate summary statistics.
\end{enumerate}
As a result of the above process, we obtain a real-valued feature matrix where the columns are derived features and the rows are unique observations. For conventional machine learning, this can be directly fed into a classification or regression model. However, each feature must be binarized for RQBoost. Several methods for create binary features were attempted in this paper. This included training simple machine learning models to just execute thresholding via an inequality. One-hot encoding through the hash trick was excluded because this would create more than a billion features, and the use of such data would not be computationally reasonable.

\section{A Note on TotalQBoost}

In this work we presented RQBoost, a implementation of QBoost~\cite{neven2009nips, neven2012qboost} that was improved by resampling. In addition to the development of QBoost, Neven et al.\ created an alternative algorithm, \textit{TotalQBoost}, that shows theoretical improvement over to QBoost but is not easily applicable to current quantum hardware due to the structure of the loss function~\cite{denchev2013binary} . For this reason, we have branched off of QBoost and not TotalQBoost to develop RQBoost.

\section{Future Work}
Controlled experimentation on the quantum annealer is necessary to better understand the relationships between solution quality and hardware parameters such as anneal time, programming thermalization time, and readout thermalization time. In the future, our work will focus on better understanding the trade-offs associated with dynamic parameter adjustment versus fixed parameters. This may involve the design of optimal controllers for the quantum annealer.

We found that RQBoost and QBoost are sensitive to initial feature space partitions when the number of features is larger than the number of possible QUBO variables. Larger, next-generation hardware with a perfect 1152 qubit graph would be capable of optimizing approximately 49 fully connected, binary variables~\cite{precision}. This means that attempting to solve QUBO problems with more than 49 variables will require approximations of the original problem. QBoost and its variants use greedy approaches to this problem.

Currently the largest possible gains from using RQBoost involve ensembling quantum methods with traditional methods. RQBoost could be used as an ensembler where the base learners are traditional machine learning algorithms. However, it is also possible to use RQBoost as a standalone algorithm and then ensemble RQBoost's results with other predictions. Even if RQBoost is not the strongest individual learning algorithm, its uniqueness may add value to a large ensemble of ML algorithms.

\makeatletter
    \clubpenalty10000
    \@clubpenalty \clubpenalty
    \widowpenalty10000
\makeatother
\bibliography{DulnyKimSubmission}

\end{document}